\newcommand{\beq}{\begin{equation}}
\newcommand{\eeq}{\end{equation}}

\documentstyle[preprint,aps,12pt]{revtex}
\begin{document}

\title{ \bf Electronic structure and ferroelectricity in SrBi$_{\bf 2}$Ta$_{\bf 2}$O$_{\bf 9}$}

\author{M.G.Stachiotti}
\address{Instituto de F\'{\i}sica Rosario, Universidad Nacional de
         Rosario, \\ 
         27 de Febrero 210 Bis, 2000 Rosario, Argentina}

\author{C.O.Rodriguez}
\address{IFLYSIB, Grupo de F\a'{\i}sica del S\a'{o}lido, C.C.565, 
 La Plata (1900), Argentina}

\author{C.Ambrosch-Draxl}
\address{Institut f\"ur Theoretische Physik, Universit\"at Graz,
 Universit\"atsplatz 5, A-8010 Graz, Austria}

\author{N.E.Christensen}
\address{Institute of Physics and Astronomy, Aarhus University,
 DK-8000 Aarhus C, Denmark}

\maketitle

\begin{abstract}
The electronic structure of SrBi$_2$Ta$_2$O$_9$ 
is investigated from first-principles, within the local density approximation,
using the full-potential linearized augmented plane wave (LAPW) method. 
The results show that, besides the large Ta(5d)-O(2p) hybridization which 
is a common feature of the ferroelectric perovskites, there is an important 
hybridization between bismuth and oxygen states. 
The underlying static potential for the ferroelectric distortion and
the primary source for ferroelectricity is 
investigated by a lattice-dynamics study using the 
frozen-phonon approach. \\ \\ 
\underline{Pacs numbers: 77.80.-e, 77.84.Dy} 

\end{abstract}

\newpage

\section{Introduction}

\noindent

Ferroelectric materials can display a wide range of dielectric, 
ferroelectric, piezoelectric, electrostrictive and pyroelectric 
properties. The potential utilization of these properties 
in a new generation of devices has motivated intensive studies.
For example, the high dielectric permittivities of perovskite-type
materials, like Sr$_x$Ba$_{1-x}$TiO$_3$, can be advantageously used 
in dynamic random access memories (DRAM), while the 
large values of switchable remanent polarization of ferroelectric materials 
are suitable for non-volatile ferroelectric random access memories 
(NVFRAM)~\cite{scott89,scott95,auc98,scott98}.     

The most popular ferroelectric materials for nonvolatile memory
applications are PbZr$_x$Ti$_{1-x}$O$_3$ (PZT), because they have 
a high Curie temperature and large remanent polarization. However, these
materials have serious fatigue degradation problems which can be solved by
the modification of the electrode. An alternative approach to control the 
fatigue problem in ferroelectric capacitors is to use other ferroelectric
materials. In recent years, SrBi$_2$Ta$_2$O$_9$ (SBT) 
has emerged
as an important candidate for non-volatile ferroelectric 
memories~\cite{des95,ara95}.
It exhibits many desirable properties
in order to be considered as an important component of the memory devices 
under development: almost no fatigue after 10$^{12}$ switching cycles, 
good retention characteristics, low switching fields and low leakage currents.

Bismuth-containing layered perovskites have been found to be ferroelectric
by Smolenskii, Isupov and Agranovskaya~\cite{smo61}. These materials belong
to the family of Aurivillius compounds with a general formula
(Bi$_2$O$_2$)$^{2+}$ (A$_{m-1}$B$_m$O$_{3m+1})^{2-}$, consisting of $m$
perovskite units sandwiched between bismuth oxide layers~\cite{aur49}
(here A and B are the two types of cations that enter the perovskite unit).

It is well known that the Aurivillius composition SrBi$_2$Ta$_2$O$_9$ (SBT)
has ferroelectric behavior at room temperature. Its crystal structure was
investigated by Rae and co-workers~\cite{rae92}. The room-temperature 
structure is orthorhombic (space group A2$_1$am) and the primitive cell 
contains 28 atoms.
The lattice parameters of the conventional unit 
cell are $a=5.531 \AA$, $b=5.534 \AA$ and $c=24.984 \AA$. 
As a result, its electrical properties are expected to exhibit 
a high degree of anisotropy. For instance, it was observed that the 
ferroelectricity along the $c$ axis is absent or very low~\cite{des96}, with
$a$ being the polar axis. More recently, this property 
was confirmed at the submiscropic level by piezoresponse scanning force
microscopy~\cite{pig99}.   

The complex crystal structure of SBT can be described in terms of relatively 
small perturbations 
from a high-symmetry body-centered tetragonal structure (space group
symmetry I4/mmm, $a = b \approx \frac{5.53}{2^{1/2}} = 3.91 \AA$), which
contains only one formula unit per primitive cell (as in other 
orthorhombic structures, the $a$ and $b$ axes are rotated 
by 45$^o$ with respect to the tetragonal case).

Although the orthorhombic symmetry is responsible for the
ferroelectricity in SBT, the parent tetragonal structure shown in Figure 1 
provides a convenient simplification for visualizing and dealing with the 
complex SBT structure. As a matter of fact, most bismuth layer compounds have 
been reported to be pseudotetragonal, with tetragonal symmetry above the 
Curie point and orthorhombic symmetry below it (for SBT the Curie temperature 
is 608K ). 
So the parent tetragonal structure of SBT could be related to the crystal 
structure of its paraelectric phase. 
Two main distortions from the tetragonal prototype structure lead to the
orthorhombic structure. First, the ions displace along the orthorhombic 
$a$ axis ([110] axis of the tetragonal structure). Second, the TaO$_6$
octahedra rotate around the $a$ and $c$ axes.  
The first factor is directly responsible for the observed macroscopic
spontaneous polarization along the $a$ direction.

In spite of its technological importance, theoretical studies on SBT are 
limited due to the complexity of its crystal structure.   
The band structure of SBT was initially calculated by Robertson et al.
using the tight-binding method~\cite{rob96}, where a highly simplified orbital
basis set was used to reproduce the main features of the bonding.
They used an orthogonal basis of O p, Ta d, and Bi s and p orbitals, and
no orbitals on the Sr. The interaction parameters were found by transferring
them from established band structures of other compounds.
Their results show that both the valence and conduction band extrema
are composed  of states localized mainly on the Bi-O layer. It was argued
that the Bi$_2$O$_2$ layer dominates the electronic response (band gap,
effective masses, etc.), while the ferroelectric response largely originates
from the SrTa$_2$O$_7$ perovskite blocks.

Several other speculations have been made about the origin of 
ferroelectricity in SBT and related compounds. Among them are the central role 
of Sr$^{+2}$ ion diplacements~\cite{lui94}, the off-center position
of Ta$^{+5}$ ion relative to its octahedron of surrounding 
oxygens~\cite{new73}, and the movement of the Ta-O plane relative to the 
Bi-O plane~\cite{rae92}.

More recently, the difference between the electronic structures  
SrBi$_2$Ta$_2$O$_9$ and SrBi$_2$Nb$_2$O$_9$ (SBN) was investigated using the 
discrete variational X-$\alpha$ cluster method~\cite{miu98}. 
The results indicate 
that the difference in the remanent polarizations in SBT and SBN is due to 
the different displacements of Ta and Nb at the B sites of a
pseudo-perovskite layer, and not due to differences in the displacement
of the other ions.  
However, one cannot expect high-accuracy results for relaxations and electronic states in crystalline materials from cluster methods. Thus there is a need for 
the application of highly precise self-consistent band structure methods.

In this work we present a first-principles study of the electronic 
structure and ferroelectric instability in SBT. The calculations were performed 
within the local density approximation to density functional theory, 
using the full-potential linearized augmented plane-wave method.

\section{Method}

The calculations presented in this work were performed 
using the full-potential linearized augmented plane-wave method (LAPW) 
method (see, e.g. Ref. \cite{sin94}) with the addition of local-orbital
basis functions~\cite{sin91} as implemented in the WIEN97 code
\cite{wien97}. Exchange and correlation effects were treated within the 
local density approximation (LDA), using the parametrization by 
Perdew and Wang~\cite{per92}.

The muffin-tin sphere radii $R_i$ = 2.0, 1.8, 2.3 
and 1.5 a.u. were used for Sr, Ta, Bi and  O, respectively. 
The value of the parameter $RK_{max}$, which controls the size of the basis
set for the wavefunctions, was chosen to be 7.3 for all the calculations. 
This resulted in well converged basis sets consisting of approximately
2500 LAPW functions. For the Sr-4$s$ and 4$p$, Ta-5$s$, 5$p$ and 4$f$, 
Bi-6$s$ and 5$d$, and O-2$s$ states local orbitals were chosen in addition.
Integrations in reciprocal space were performed using the tetrahedron
method. We used a 6$\times$6$\times$6 mesh which represents 
28 k-points in the irreducible wedge for the body-centered tetragonal 
structure.
Convergence tests indicate that only small changes result from 
going to a denser k-mesh or to a larger value of $RK_{max}$.

\section{Results and discussion}

\subsection{Electronic structure}

The electronic structure of SBT is calculated for the tetragonal 
parent structure. We use the experimental lattice constants   
$a = 3.91$ \AA~and $c = 24.984$ \AA. Since the internal parameters of this
structure have
not been determined experimentally, we evaluated the equilibrium
positions of the atoms using a damped Newton dynamics method. (The 
final force on each atom was less than $\approx$ 1 mRy/a.u.)  
The equilibrium coordinates are listed in Table I. 

The band structure of the optimized geometry is shown in Figure 2 
along several high-symmetry lines in the Brillouin zone. 
Roughly speaking, the two bands centered at 
$\approx -10$ eV are derived from Bi 6s orbitals while the manifold 
of 27 valence bands are derived mainly from O 2p orbitals. 
We found that the fundamental band gap is indirect, since the valence band
maximun lies at $X$ while the conduction band minimun is the Brillouin zone center $\Gamma$.
As is typical in LDA calculations for semiconductors, the band gap 
is underestimated. The experimental band gap obtained from UV-absorbance 
measurements is 4.2$\pm$0.2 eV~\cite{har97}, which is twice as large than our
theoretical value of  $\approx $ 2 eV.
It is interesting to note that, in disagreement with our results, 
an indirect ($\Gamma$-M) band gap was obtained by the tight-binding 
calculation~\cite{rob96}, which indicates important differences in the 
details of the dispersion. 

The total density of states between -35 and 5 eV is shown in Figure 3 
illustrating the energy position of the semicore states.   
The calculated total DOS is in reasonable agreement with XPS data, which show 
a valence band width of $\approx$ 7 eV and a peak centered $\approx$ 10 eV 
below the valence band maximum (Bi 6s band)~\cite{har97}.

Although the valence band presents mainly O 2p character, there is a quite
strong hybridization with Bi and Ta states, as is evident from the 
sphere-projected density of states (DOS) shown in Figure 4. Examination of 
the DOS reveals that there is substantial O 2p character in the conduction 
bands,
rising from zero at the conduction band minimun with 
increasing energy.
Conversely, there are strong Ta and Bi contributions to the valence band. 
The Ta 5d contribution is zero at the valence band maximun but rises strongly
with decreasing binding energy, reflecting the Ta 5d-O 2p covalency (as it 
is common in ABO$_3$ perovskites). The Bi contribution has both s-like
(mainly in the upper part of the valence band) and p-like  
(mainly in the lower part) character. 
These contributions arise 
from a strong hybridization between O(2,3) 2p valence band states with 
Bi 6s (fully occupied) and Bi 6p (conduction) bands.  
A detalied examination of the band characters reveals that 
the valence band maximun and the conduction band minimun are not mainly
localized in the Bi-O(3) layer, as was obtained by the tight-binding
calculation~\cite{rob96}. In fact, in our case the valence band  maximun 
at X is primarily of O(1,4) 2p character.

We finally show in Figure 5 the valence charge density map of SBT in two 
high symmetry planes, where the Ta-O and Bi-O "covalent bonds" can be
seen.

\subsection{Ferroelectric instability}

The main features of  
SBT described so far resembles the electronic structure of PbTiO$_3$.
In this material, besides the large Ti 3d-O 2p hybridization,
there is an important hybridization between 
Pb 6s and O 2p states; and this covalent bond plays a 
central role in the stabilization of the tetragonal ferroelectric 
structure of PbTiO$_3$~\cite{coh92}. 

The presence of quite strong Ta-O and Bi-O hybridizations in SBT
opens fundamental questions about the origin of its ferroelectricity:
What is the underlying static potential for the ferroelectric distortion?
Which is the primary source for ferroelectricity? 
We investigate these questions by a lattice dynamics study using the 
frozen-phonon approach.

To search for the presence of a possible lattice instability of
the tetragonal structure, we determined the phonon frequencies and 
eigenvectors of the infrared-active $E_{\mu}$ modes,
which are polarized perpendicular to the $c$ axis.
To this end, we calculated atomic forces for several small displacements
($\sim 0.01$~{\AA}) consistent with the symmetry of the mode.
From the force as a function of displacement,
the dynamical matrix was constructed and diagonalized.
The calculated frequencies and eigenvectors are listed in Table II.

While experimental studies were carried out on infrared-active (IR) phonons
in orthorhombic SBT, using reflectivity and transmission 
measurements~\cite{mor98}, 
there is no experimental determination of IR phonon
frequencies for the tetragonal structure to directly compare with. 
The remarkable fact of our calculation is, however, the presence of one 
unstable phonon mode. 
Roughly speaking, this mode mainly involves movements of the Bi atoms 
with respect to rest of the lattice. Actually, its displacement pattern is 
(0.085, 0.114, -0.206, 0.325, 0.549, 0.053, 0.224, 0.182), showing 
large displacements of the TaO$_6$ perovskite-like blocks relative to 
Bi atoms. This displacement vector is obtained from the 
eigenvector by dividing each component by the square root of the 
corresponding atomic mass and then normalizing to unity.
So, the polarity of the mode can be described as a vibration 
of the TaO$_6$ perovskite-like block relative to Bi, with an additional 
contribution arising from the movement of Ta relative to its surrounding 
oxygens.
Rae et al.~\cite{rae92} argued that the polarity of orthorhombic SBT can be 
quite well described as a movement of Ta, O(4) and O(5) relative to Bi 
and O(3) of the BiO$_2$ layer with and additional movement of Ta relative
to O(4) and O(5). Although the displacement pattern of the unstable mode is
not directly related with the atomic positions of the orthorhombic 
phase, both pictures show some consistence.  
It is worth to mention that, from a detailed analysis of the calculated 
force constants, the lattice instability primarily arises from the 
attractive Bi-O(2) interaction. 

The resulting in-phase movement of the Ta atoms with respect to their
oxygen octahedra (obtained from the eigenvector of the ferroelectric mode), 
could be the explanation of the low remanent polarization observed in SBT. 
However, this situation could be different for the case of SBN, considering  
the different structural behavior of KNbO$_3$ and KTaO$_3$.
While KNbO$_3$ has a series of ferroelectric phase transitions, KTaO$_3$
remains cubic down to low temperatures. Furthermore, it was emphasized 
by Singh~\cite{sin96}, that the absence of ferroelectricity in KTaO$_3$ is 
due to the extreme sensitivy of the soft-mode to the covalency and the slight
chemical differences of Nb and Ta, particularly the higher d binding energy of 
Nb. This point will be clarified in the future by a corresponding investigation 
on SBN. 

Finally, the total energy is evaluated as a function of the relative 
displacement
of Bi with respect to O(2) corresponding to the unstable mode. The results are shown in Figure 6 which depicts the energy per formula unit for displacement patterns along the [100] and [110] direction, respectively. Ferroelectric instabilities with energy gains of $\approx$ 4 mRy/cell and 6 mRy/cell for 
the [100] and [110] directions are observed. Although the ferroelectric
mode mainly involves displacements of the Bi atoms 
with respect to rest of the lattice, a [110] displacement of the Bi sublattice  
alone does not produce a lattice instability (see Figure 6). 
This indicates that Bi has not a tendency to go off-center in the tetragonal 
phase, and the energy wells presented in the figure are indeed associated
to the specific pattern of atomic displacements of the ferroelectric mode.

As already mentioned, the bonding of these materials
stems from the transition metal-oxygen hybridization. As can be seen from
Figure 7, the whole spectrum of eigenvalues gets modified in an
intricate way upon the complex distortion that gives rise to the
instability. The two main effects are an increase in the band
gap and a decrease of the separation between the Bi-s states
and the valence band. When the distortion sets in, spectral weight from the
lower part of the conduction band gets reduced
and transfered to higher energies. This is more 
pronounced in the Bi contribution to the DOS than in that of Ta.
In the valence region, the DOS gets reduced at the top 
and increased at the bottom. Here, the effect on the Ta contribution 
is stronger.
So, both covalent bonds seem to play an important role. It is worth to mention
however that the energetics shown in Figure 6 finally comes from a very 
delicate balance between several contributions: covalent, Coulomb and
repulsive ionic interactions.

The [110] displacement results in a deeper total energy minimun, which is
consistent with the fact that the low temperature ferroelectric state of SBT 
has orthorhombic symmetry. The existence of a saddle point in the [100] 
direction on the total energy surface could indicate that the phase 
transition in SBT is not as simple as considered until now.
In order to discuss this point it is useful to remark some aspects of the
dynamical mechanism leading to the sequence of phase transitions in ABO$_3$ 
perovskites. Theses crystals have been considered for a long time as 
displacive-type ferroelectrics. The main evidence for this behavior has been 
the existence of a $\Gamma$-TO soft mode which has been observed in many
perovskites \cite{coc60,sco74}.
However, there is experimental \cite{fon84,vog86,fon88,sok88,fre97} 
as well as theoretical~\cite{pos93,sep97,sta98} evidence that the 
sucessive phase transitions in KNbO$_3$, and BaTiO$_3$, have pronounced 
order-disorder features, and that they are quite well described in the 
framework of the eight-site model. 
According to this order-disorder approach, the total energy surface 
has a maximum for the cubic perovskite structure, eight degenerate minima
for the [111] soft mode amplitude displacements, and saddle points for the 
[100] and [110] displacements.
In the cubic phase, the eight minima are occupied with equal probability, 
where this symmetry is broken as the temperature is lowered: Four sites are 
occupied in the tetragonal phase, two sites in the orthorhombic phase, and 
finally, only one site is occupied in the rhombohedral structure.
In this way, the relaxation process which displays a critical slowing down
when T$_c$ is approached in the different phases is interpreted to be the
driving mechanism of the phase transitions.

The lattice instability related to the ferroelectric phase transition
of SBT was recently studied by Raman scattering in the temperature range 
between 293K and 958K~\cite{koj98}. 
The lowest frequency mode of 29 cm$^{-1}$ at room temperature showed
remarkable temperature variations towards T$_c$. While its frequency
decreases markedly, the damping factor increases rapidly below T$_c$. This fact
may suggest that the nature of the phase transition shows a crossing over from
displacive to order-disorder type in the neighbourhood of T$_c$.
As in KNbO$_3$, it was also found, that the extrapolated frequency is still 
finite at T$_c$, where this relatively large value may originate not only from 
the first order phase transition, but also due to the coupling between the 
soft mode and the strain.     

If this coupling between soft mode and strain is strong enough, 
the energetics shown in Figure 6  would give rise to the presence of an
intermediate phase in SBT, leading to the following phase transition sequence:
paraelectric (tetragonal) $\rightarrow$ ferroelectric (net polarization along
the [100] direction $\rightarrow$ ferroelectric (orthorhombic, net polarization
along the [110] direction).
As in the eight-site model, the four energy minima would be occupied with 
equal probability in the high-temperature paraelectric phase, two sites 
would be occupied in the intermediate phase (with a net polarization
along the [100] direction), and finally, one site would be occupied in the 
orthorhombic phase.  

Recently, an anomaly at T=520K was observed in specific heat 
measurements of SBT films~\cite{ono99}. For Bi-rich 
Sr$_{0.8}$Bi$_{2.2}$Ta$_2$O$_9$ two anomalies were found at 620K and 410K, with 
corresponding changes in the X-ray diffraction pattern suggesting structural
changes. The introduction of a small amount of excess Bi improved significantly 
the ferroelectric properties of SBT (the spontaneous polarization is two times
larger and the Curie temperature shift to 670K from 608K~\cite{tak98}). 
These results also suggest that the phase transition in SBT is not as 
simple as considered until now. More detailed examinations of crystal
symmetry and dielectric measurements on single crystals are necessary for
clarifying the nature of ferroelectricity in the SBT family compounds.      


\acknowledgments
M.G.S. thanks CONICET, CIUNR and FONCyT for support. CAD acknowledges support
from the Austrian Science Fund, project P13430-PHY.

\newpage

\begin{center}
{\bf {\large Figure Captions}}
\end{center}

{\bf Figure 1}: \\
Tetragonal structure of SrBi$_2$Ta$_2$O$_9$. 
Only atoms between $\frac{1}{4}c$ and  $\frac{3}{4}c$ are shown. \\ \\

{\bf Figure 2}: \\
Band structure of SrBi$_2$Ta$_2$O$_9$
along several high-symmetry lines in the Brillouin zone.\\ \\

{\bf Figure 3}: \\
Total density of states for SrBi$_2$Ta$_2$O$_9$, between -35 and 5 eV, showing
the energy position of the semicore states. \\ \\

{\bf Figure 4}: \\
Total and site-projected electronic densities of states for the valence 
and conduction bands of SrBi$_2$Ta$_2$O$_9$. \\ \\

{\bf Figure 5}: \\
Valence charge density plot in the (100) and (110) planes of 
SrBi$_2$Ta$_2$O$_9$. The scale is from 0 to 0.2 electrons per bohr$^3$
and the contour interval is 0.01 electrons/bohr$^3$. \\ \\

{\bf Figure 6}: \\
Energy as a function of the distortion along the [110] ($\bullet$) and [100]
($\Box$) directions corresponding to the unstable mode. 
The normal coordinate is represented by a motion of Bi relative to O(2). 
The energies are with respect to that of the perfect tetragonal structure.
The energy as a function of the [110]-displacement of the Bi sublattice  
is also shown ($\triangle$). \\ \\

{\bf Figure 7}: \\
Comparison of the partial density of states for Ta and Bi 
in the tetragonal structure, with (full line) and without (dotted line) 
displacements of the ions, according to the displacement pattern of the unstable-mode along the [110] direction.

\newpage

\begin{table} 
\vspace*{2.0 truecm}
\label{table1}
\caption{Equilibrium atomic coordinates in lattice constant units (a=3.91$\AA$ 
and c=24.984 $\AA$).} 
\vspace*{0.3cm}         
\begin{tabular}{|c|ccc|} 
  atom     & X & Y  & Z \\ \hline
Sr         & 0.0 & 0.0  & 0.0  \\
Ta         & 0.5 & 0.5  & $\pm$ 0.08468  \\
Bi         & 0.0 & 0.0  & $\pm$ 0.20124 \\
O(1)       & 0.5 & 0.5  & 0.0  \\
O(2)       & 0.5 & 0.5  & $\pm$ 0.16033  \\
O(3)       & 0.0 (0.5) & 0.5 (0.0) & 0.25  \\
O(4)       & 0.0 & 0.5  & $\pm$ 0.07602  \\
O(5)       & 0.5 & 0.0  & $\pm$ 0.07602 \\
\end{tabular}
\end{table}

\begin{table}
\vspace*{2.0 truecm}

\caption{Frequencies $\omega$ and eigenvectors of the $E_{\mu}$ modes, which are
polarized perpendicular to the $c$ axis. } 
\vspace*{0.3cm}         
\label{table2}
\begin{tabular}{c|cccccccc} 
 $\omega$  &  \multicolumn{8}{c}{Eigenvector}   \\      
  (cm$^{-1})$   & Sr    & Ta+   & Bi+    & O1    & O2 +  & O3+   & O4+     & O5+
  \\ 
  \hline
 495 & 0.036 & 0.052 & 0.001  & 0.432 & 0.033  & -0.001 & -0.617 & 0.148 \\
 247 & -0.051 & 0.064 & 0.053  & -0.754 & 0.324  & -0.205 & -0.212 & 0.125 \\
 215 & -0.032 & 0.053 & -0.117  & -0.360 & -0.122  & 0.621 & -0.113 & 0.073 \\
 113 & -0.786 & 0.395 & -0.026  & 0.019 & -0.120  & -0.066 & -0.019 & -0.122 \\
 100 & 0.147 & 0.037 & 0.031  & 0.010 & 0.239  & 0.108 & -0.129 & -0.634 \\
 73  & 0.502 & 0.301 & -0.191  & -0.232 & -0.403  & -0.199 & -0.088 & -0.103 \\
 i54 & 0.130 & 0.251 & -0.489  & 0.213 & 0.360  & 0.035 & 0.147 & 0.120 \\
\end{tabular}
\end{table}

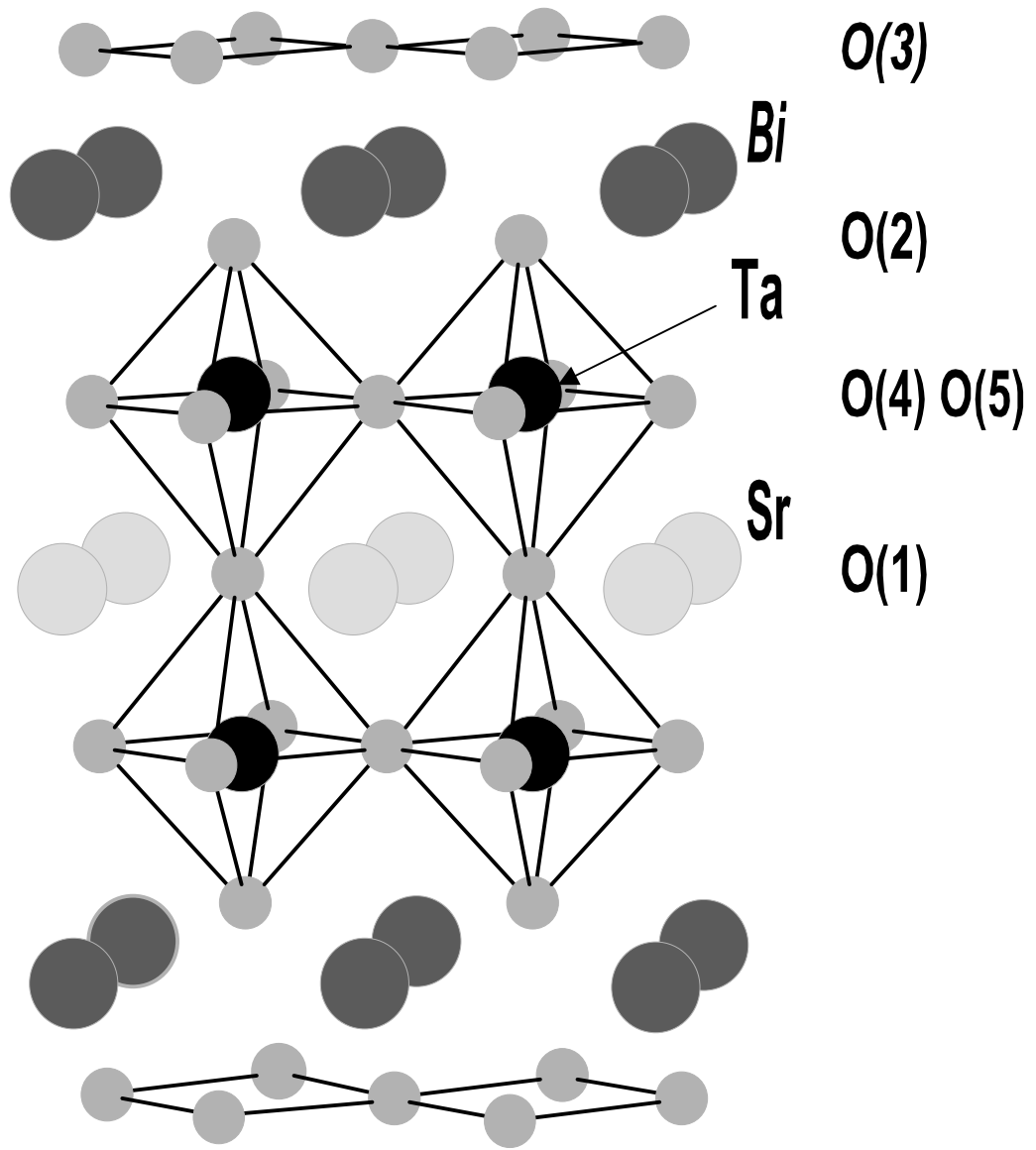
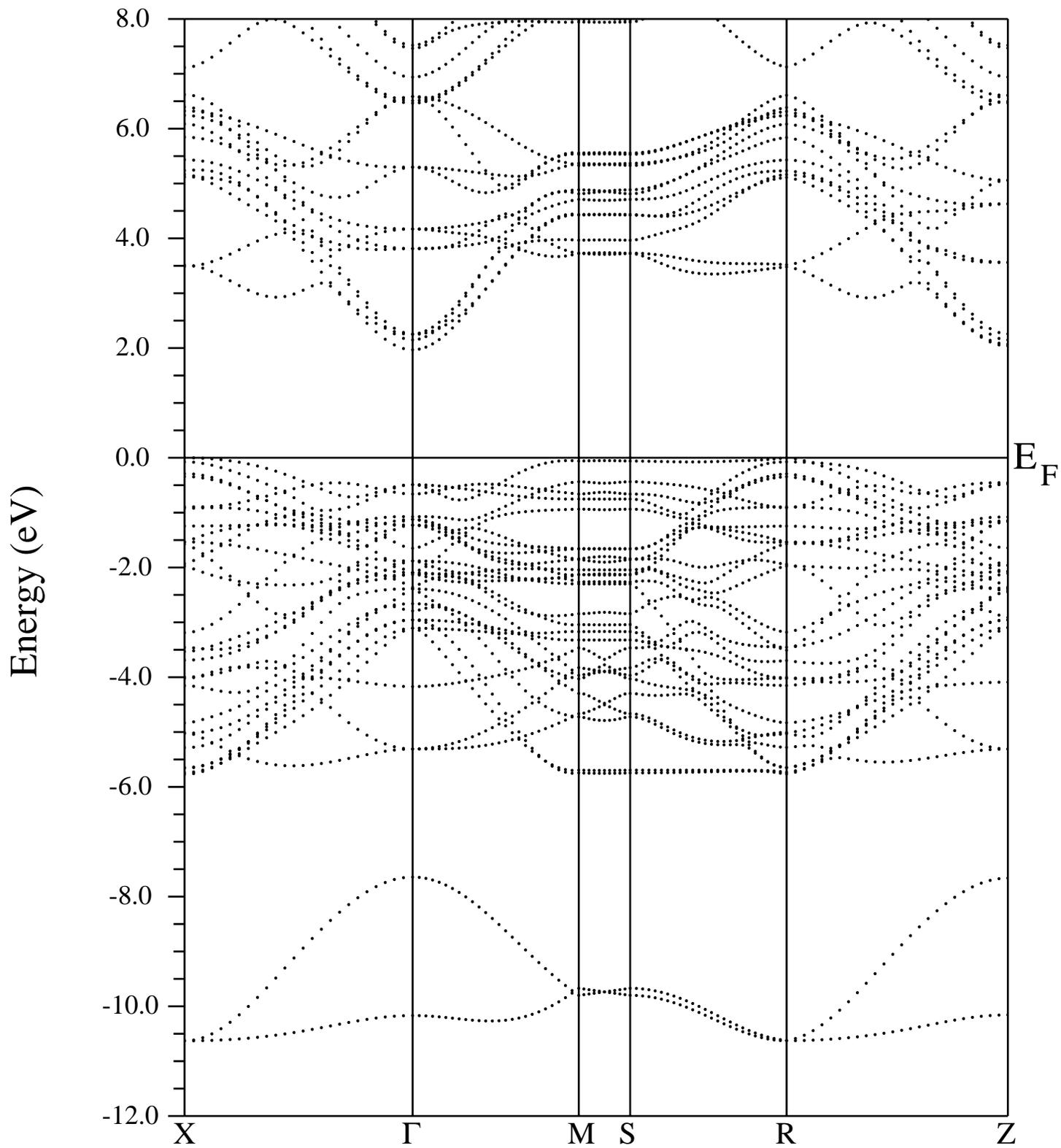
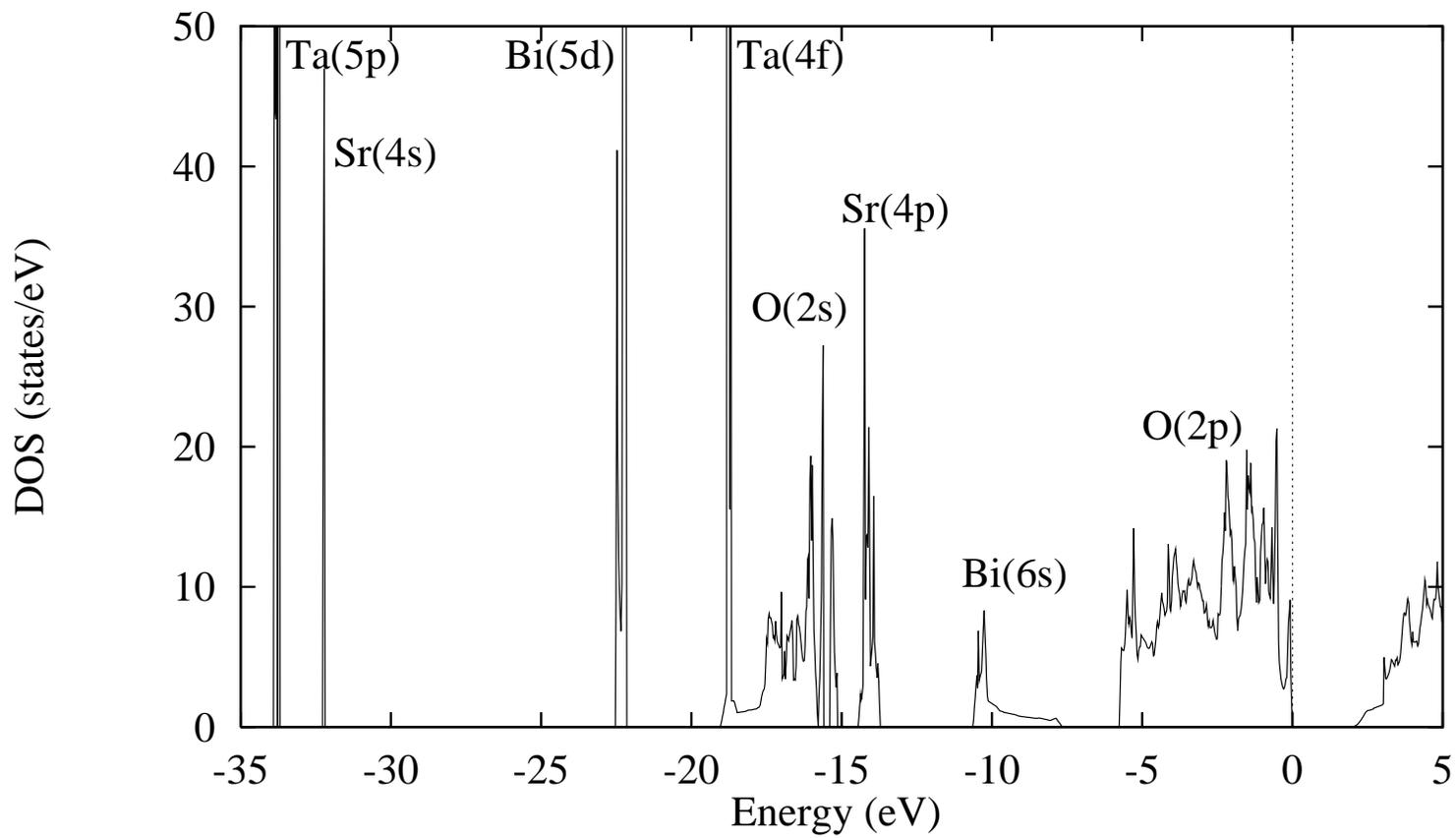
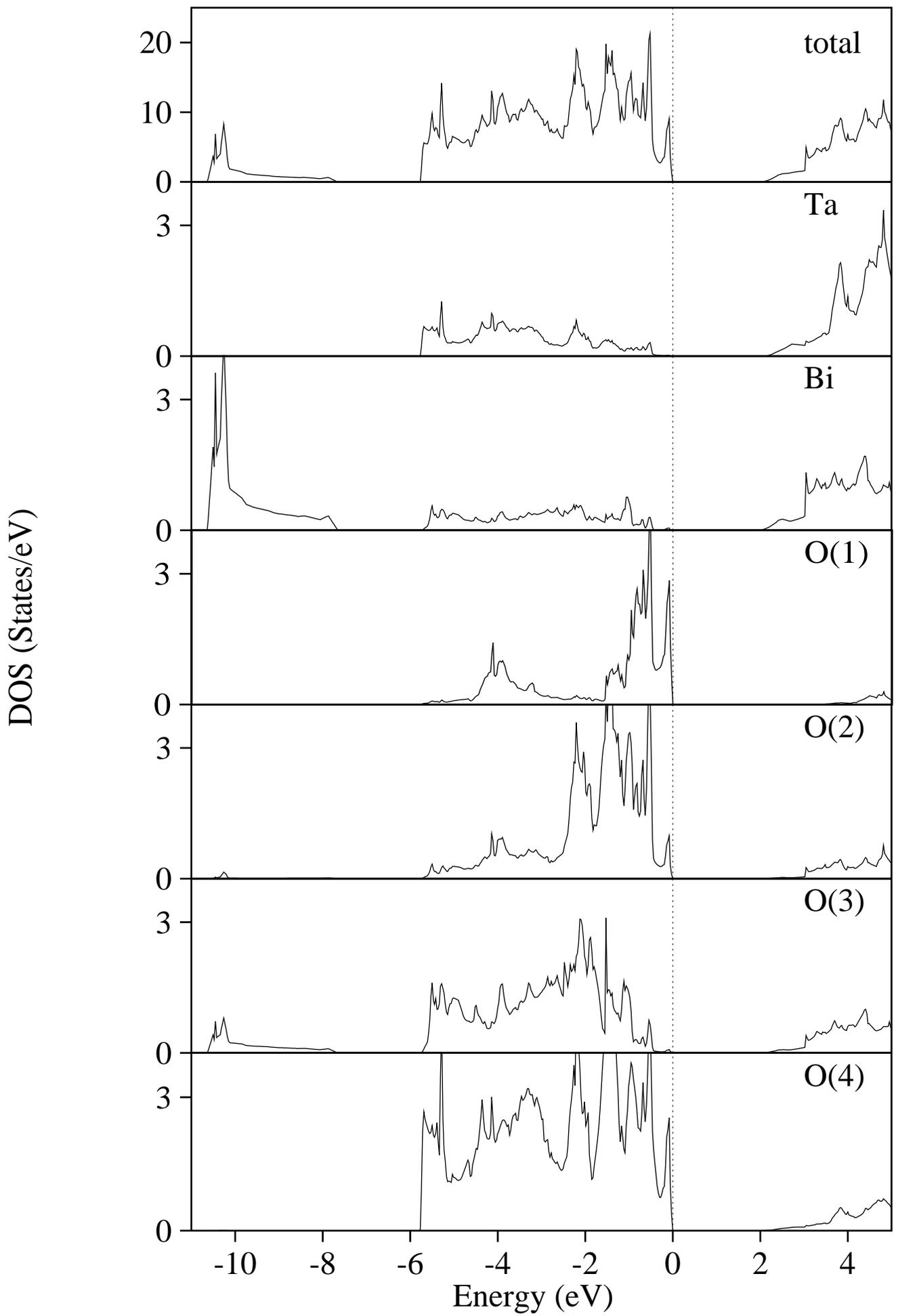
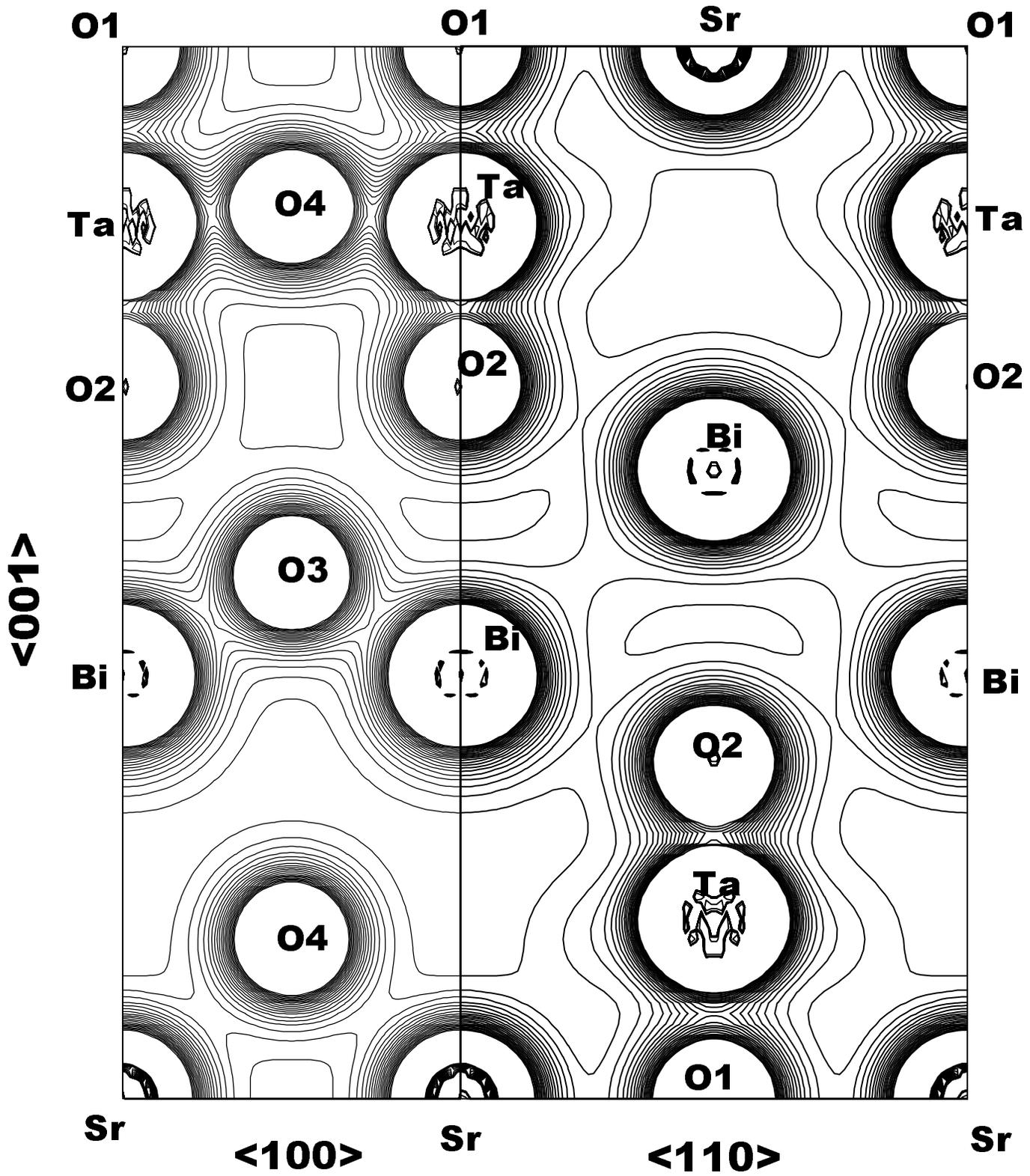
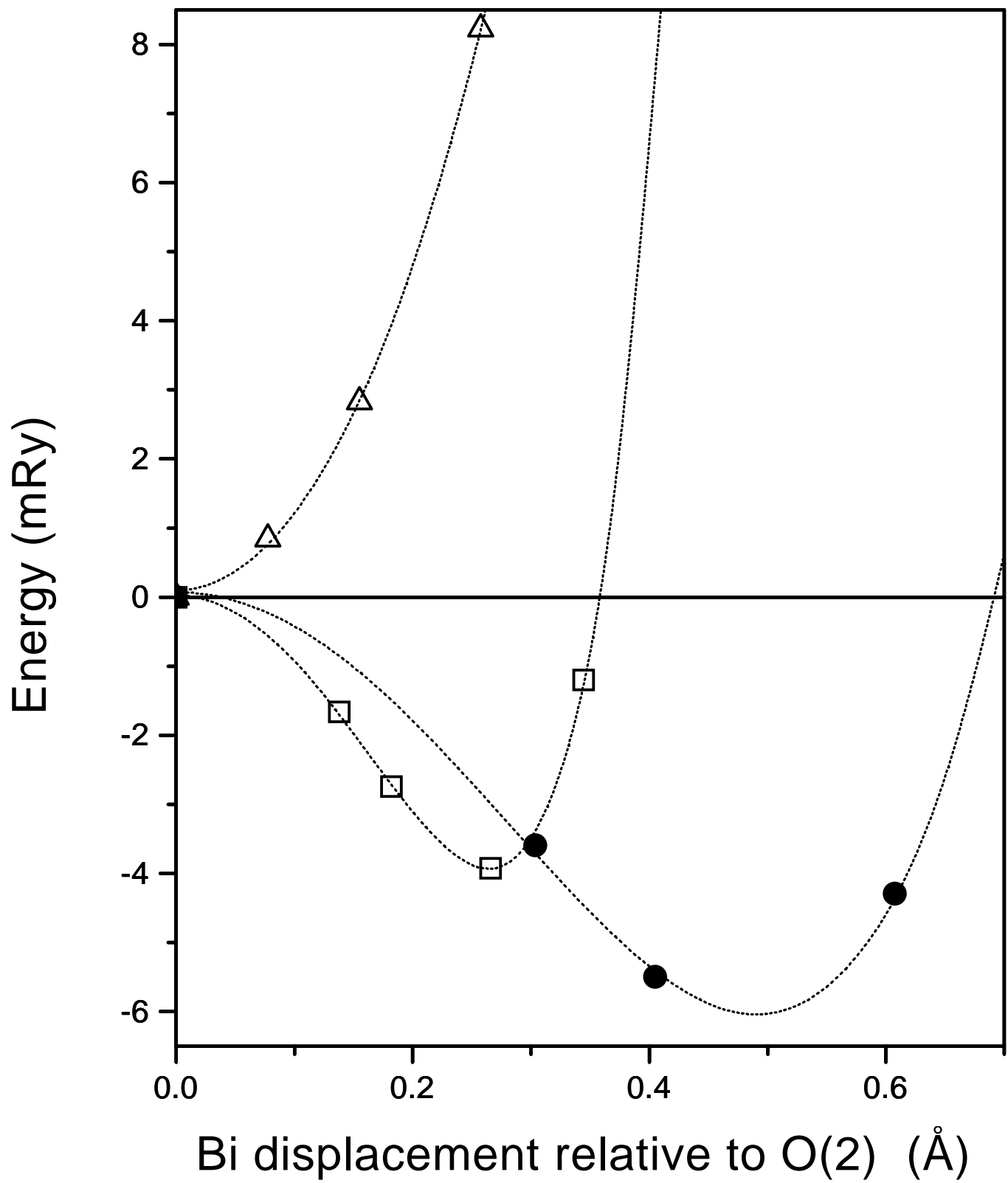
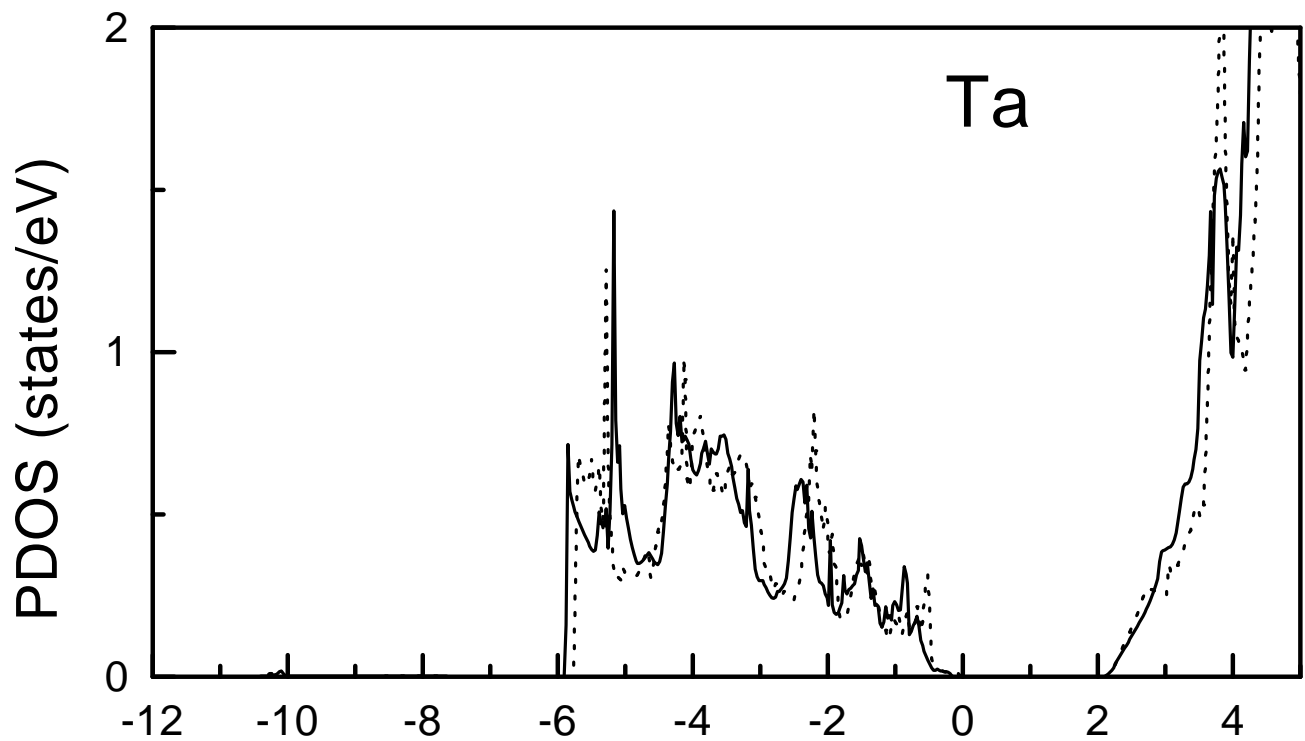
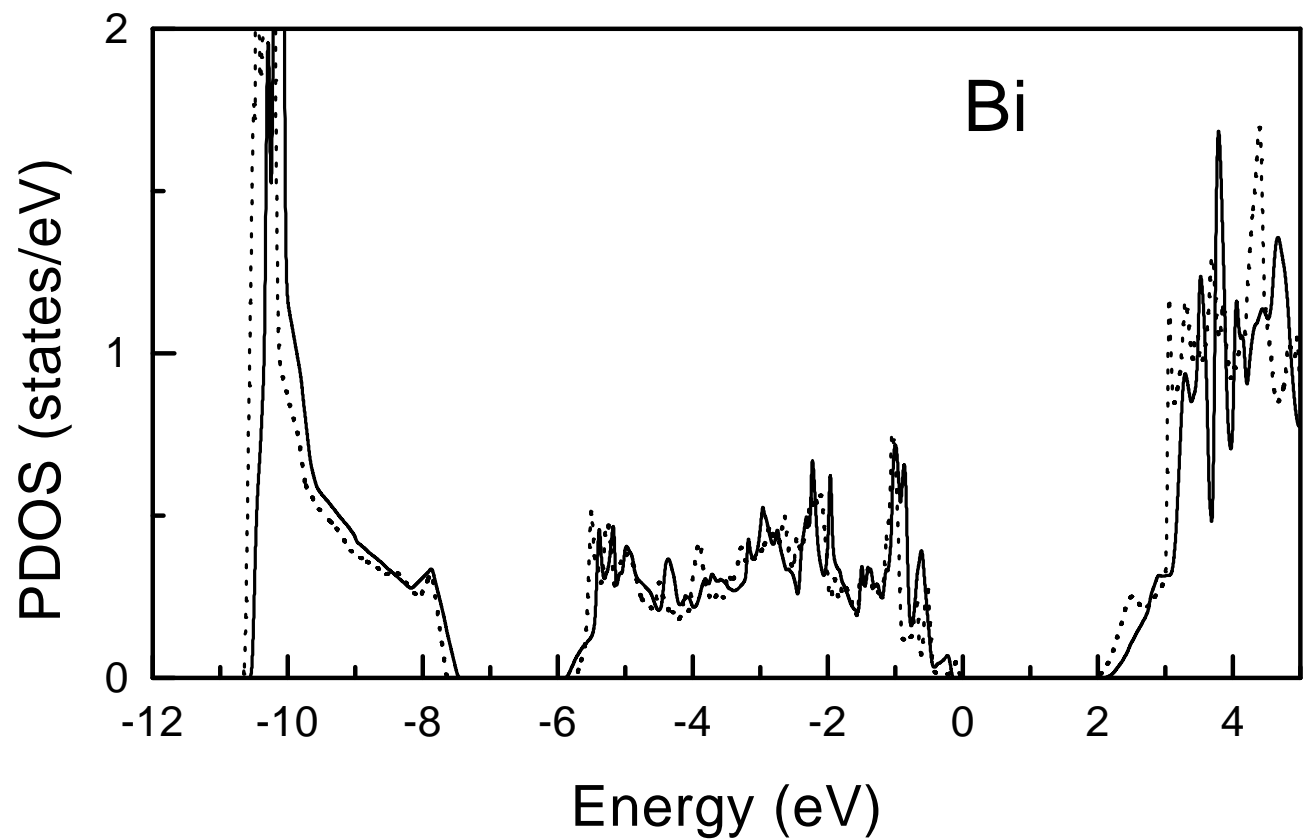
\end{document}